\documentclass[12pt]{article}
\let\section=\subsection     \let\subsection=\subsubsection

\usepackage{graphicx,epsfig}

\begin{document}
\begin{center}
   {\large \bf SCALING PROPERTIES OF THE TRANSVERSE MASS SPECTRA}\\[2mm]
   J.~SCHAFFNER-BIELICH$^a$, D.~KHARZEEV$^b$, L.~MCLERRAN$^b$  and 
  R.~VENUGOPALAN$^{b,c}$ \\[5mm]
   {\small \it  
$^a$Department of Physics, Columbia University, New York, NY 10027, USA\\ 
$^b$Department of Physics, Brookhaven National Laboratory, 
Upton, NY 11973, USA\\ 
$^c$RIKEN BNL Research Center, Brookhaven National Laboratory,\\  
Upton, NY 11973, USA\\[8mm] }

\end{center}

\begin{abstract}\noindent
Motivated from the formation of an initial state of gluon-saturated matter, we
discuss scaling relations for the transverse mass spectra at BNL's
Relativistic Heavy-Ion Collider (RHIC). We show on linear plots, that the
transverse mass spectra for various hadrons can be described by an universal
function in $m_t$. The transverse mass spectra for different centralities can
be rescaled into each other. Finally, we demonstrate that $m_t$-scaling is
also present in proton-antiproton collider data and compare it to
$m_t$-scaling at RHIC.
\end{abstract}

% add discussion about radial flow in ppbar collisions of Peter and Berndt!!!

In this work, we report on applying scaling features of the initial state of the color
glass condensate to data at RHIC, in particular to the transverse
momentum distribution of hadrons (see \cite{mlsb01,skmv02}).

It has been suggested that in collisions of heavy-ions at ultrarelativistic
bombarding energies gluons are so densely packed in phase-space that they
saturate \cite{Gribov} and form a color glass condensate
\cite{mv,kmw,kov,jkmw,iancu00,kn}.  The initial momentum distribution of
gluons for a color-glass condensate can be described by the scaling relation
\cite{alex00}:
\begin{equation}
        {1 \over \sigma} {{dN_g} \over {d\eta d^2p_t}} 
                        = \frac{1}{\alpha_s(Q_s^2)} 
        f_g \left(\frac{p_t^2}{Q_s^2}\right) \quad .
\label{eq:scaling_charged}
\end{equation} 
Here, $\sigma$ is the transverse area, $Q_s$ the saturation scale and $f_g$ an
universal dimensionless function for the produced gluons.  The saturation
scale $Q_s$ depends on the energy, the centrality and atomic number of the two
colliding nuclei.

We have to make several assumptions about the underlying dynamical picture to
compare the above ansatz to actual experimental data: the initial gluon
distribution is characterized by saturated gluons, there is a free streaming
evolution which produces additional partons and finally, there is a freeze-out
to hadrons.  At all stages, the scaling properties of the initial state are
preserved. Hence, we replace the scaling function for gluons and the
saturation momentum with those for hadrons, i.e.\ $f_g\to f$ and $Q_s \to
p_s$, respectively.  Note, that $p_s$ is not equal to the saturation momentum
for gluons but should have similar dependencies on energy and centrality. This
simple picture works to explain the centrality dependence and the
pseudorapidity dependence of the charged multiplicity at RHIC \cite{kn}. The
question we are asking is, whether the above scaling relation
eq.~(\ref{eq:scaling_charged}) can be utilized to describe the transverse
momentum spectra at RHIC. We point out that our assumptions are completely
contrary to the notion of radial flow or of thermalized and equilibrated
matter.

Guided by the scaling relation for saturated gluons we propose that
the distribution of produced hadrons can be cast in the following form:
\begin{equation}
        {1 \over \sigma} {{dN_h} \over {dyd^2m_t}} = \frac{1}{\alpha_s(p_s)} 
        \kappa_h \cdot f \left(\frac{m_t}{p_s}\right) 
\quad .
\label{eq:scaling_mt}
\end{equation}
The appropriate replacements, $f_g\to f$ and $Q_s\to p_s$ are made from
eq.~(\ref{eq:scaling_charged}) to eq.~(\ref{eq:scaling_mt}). In addition, we assume
that the transverse momentum has to be replaced by the transverse mass for
identified hadrons, in absence of any other boost invariant quantity.  The new
factors $\kappa_h$ take into account effects of conserved quantum numbers, as
strangeness and baryon number and are constants. The parameters $\sigma$ and
$p_s$ have to be determined from the data.

We take the minimum bias data of the transverse mass spectra from the PHENIX
collaboration as measured for gold-gold collisions at $\sqrt{s}=130$ GeV
\cite{Julia01} and plot it as a function of the transverse mass. The curves for
different hadron species are very close to each other without any adjustments
in the absolute normalization.  The spectra is closer to a power law than to an
exponential. Shifting the protons up a factor $\kappa_h=2$
(there is definitely a finite baryon number at midrapidity) and the kaons down
by a factor $\kappa_h=1/2$ (there seems to be a suppression of strange
particles) puts all curves on top of each other (see \cite{skmv02} for the
corresponding plots). That means that the slope of the curve is the same for a
given $m_t$ irrespective of the hadron. The $\kappa_h$ factors are indeed
constants, they do not depend on $m_t$. The scaling pattern is present even for
quite large values of $m_t= 3$--4 GeV. 

\begin{figure}[t]
  \centerline{\epsfig{figure=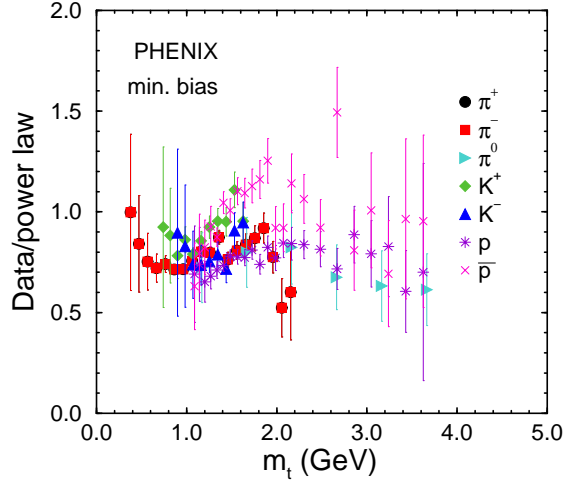,height=0.33\textheight}}
\caption{Transverse mass spectra of identified hadrons for minimum bias
  gold-gold collisions as measured for $\sqrt{s}=130$ AGeV at RHIC (preliminary
  data taken from \protect\cite{Julia01,GaborQM}). The data is normalized to a
  power law fit. The proton and kaon spectra are rescaled by a factor of 1/2
  and 2, respectively.}
\label{fig:mt_minbias}
\end{figure}

To better see any deviations from $m_t$-scaling, we divide the data by a power
law fit $\sim 1/(p_s+m_t)^n$ with $p_s=2.71$ GeV and $n=16.3$. The value for
$n$ seems to be high.  Nevertheless, we remind the reader that the two
parameters of the power law fit are tightly constrained by the mean $p_t=2
p_s/(n-3)$, so that lower values of $n$ can be achieved by lowering $p_s$
correspondingly with basically the same fit quality. The linear plot of the
transverse mass spectra relative to that power law fit is shown in
Figure~\ref{fig:mt_minbias}. One sees, that the deviations from scaling are
only about 30\% including systematic errors from the fit!

\begin{figure}[t]
  \centerline{\epsfig{figure=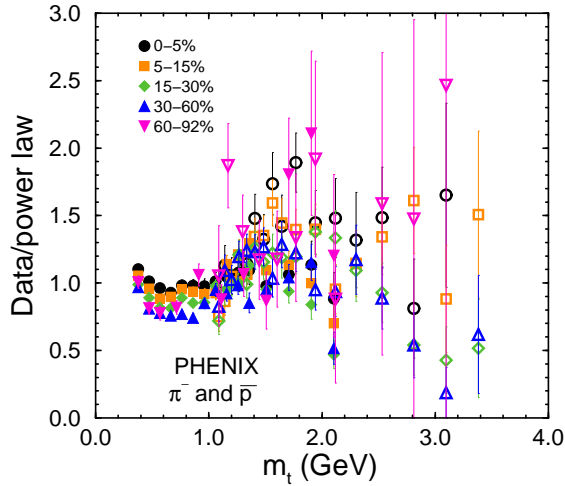,height=0.33\textheight}}
\caption{Transverse mass spectra of $\pi^-$ and $\overline{\rm p}$
  for different centralities at RHIC (preliminary data as published in
  \protect\cite{Julia01}).  Filled symbols are for $\pi^-$, open symbols for
  antiprotons. The data is normalized to a power law fit.}
\label{fig:mt_centrality}
\end{figure}

Next, we consider the centrality dependence of the particle spectra using the
transverse momentum distributions of $\pi^-$ and antiprotons from PHENIX
\cite{Julia01}. If there is indeed an universal function, the observed
$m_t$-scaling should also hold for different centrality bins. In addition the
centrality dependence is controlled solely by the transverse area $\sigma$ and
the saturation momentum $p_s$. To test the universality, the distribution for
different centralities is rescaled by
\begin{equation}
  {1 \over \sigma} {{dN_h} \over {dyd^2m_t}} \to \frac{1}{\lambda} {1 \over
    \sigma} {{dN_h} \over {dyd^2m_t}} \qquad \mbox{ and } \qquad m_t \to
  \frac{m_t}{\lambda'} \quad.
\end{equation} 
The data points are then on top of each for all centrality bins (see
\cite{skmv02} for a plot).  Again, we make a linear plot by dividing the
rescaled data by a power law. We take as reference curve a fit to the most
central bin with the parameters $p_s=1.65$ GeV and $n=11.8$. The data relative
to this power law is plotted in Figure~\ref{fig:mt_centrality}. As one sees,
scaling works within 30\% up to say $m_t\sim 2$ GeV for all centrality
classes.

The extracted scaling parameters $p_s$ and $\sigma$ are now expected to scale
like $N_{\rm part}^{1/6}$ and $N_{\rm part}^{2/3}$, respectively. The
saturation scale $p_s$ turns out to change as a function of centrality like
$p_s^2/p^2_{s,c}=0.61+0.39(N_{\rm part}/347)^{1/3}$. This implies that there is
an additional constant scale involved which is already present in pp or
p$\bar{\rm p}$ collisions and can be associated with the Hagedorn temperature
or equivalently the constant and finite mean transverse momentum seen at the
ISR of about 300 MeV. The additional vacuum scale is compatible with the mean
$p_t$ measured by the UA1 collaboration for p$\bar{\rm p}$ collisions at
$\sqrt{s}=200$ GeV \cite{ua1} of $\langle p_t \rangle = 392 \pm 3 $ MeV and the
one by the STAR collaboration of $\langle p_t \rangle =
508\pm 12$ MeV for central gold-gold collisions at $\sqrt{s}=130$ GeV
\cite{star}. The other scaling parameter, $\sigma/\alpha_s(p_s)$,
follows the expected pattern, i.e.\ $\sigma/\alpha_s(p_s)\sim N_{\rm
part}^{2/3}\ln\left(p_s^2/\Lambda_{\rm QCD}^2\right)$ \cite{skmv02}.

\begin{figure}[t]
  \centerline{\epsfig{figure=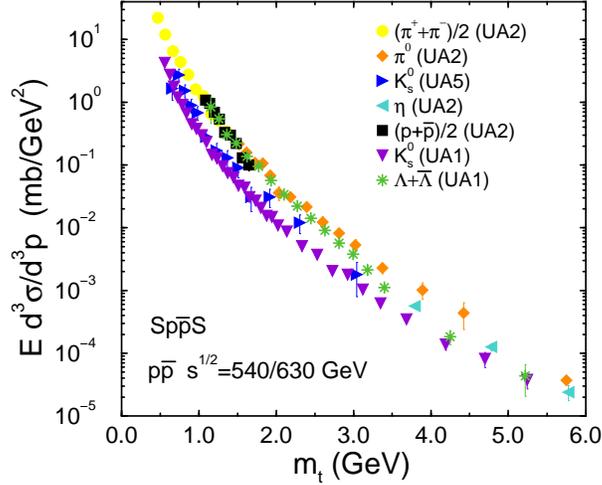,height=0.33\textheight}}
\caption{Transverse mass spectra of identified hadrons from p$\bar p$
  collisions at $\sqrt{s}=630$ GeV (UA1 data from \protect\cite{ua1_mt}) and at
  $\sqrt{s}=540$ (UA2 and UA5 data from \protect\cite{ua2_mt,ua5_mt}).}
\label{fig:mt_ppbar}
\end{figure}

Finally, we discuss $m_t$-scaling for p$\bar{\rm p}$-collisions. Identified
hadron spectra were measured for energies of $\sqrt{s}=540$ and $\sqrt{s}=630$
GeV by the UA1, UA2 and UA5 collaborations \cite{ua1_mt,ua2_mt,ua5_mt} over a
large range of transverse momenta. The transverse mass spectra of various
hadrons is depicted in Figure~\ref{fig:mt_ppbar}. Without any change in the
normalization, all curves are close to each other and have similar shapes. The
proton data (black squares) are on top of the charged pion data, i.e.\ the
$\kappa_h$ is one for the case of p$\bar{\rm p}$ collisions as the net baryon
number is zero. Kaons are suppressed compared to pions and protons, similarly
as seen in the RHIC data. The strange $\Lambda+\bar\Lambda$ data points are on
the pion and proton data, but note that there is a factor two difference
compared to the proton data, so that $\Lambda$'s are suppressed by 1/2 like the
kaons. We expect then a similar pattern for the hyperon data in heavy-ion
collisions in RHIC, if corrected for the baryon number. So, $m_t$-scaling
seems to work also for p$\bar{\rm p}$ collisions up to several GeV in
transverse mass. The $\kappa_h$ factors, i.e.\ corrections to scaling due to
strangeness and baryon number, follow the expected patterns.  One notes,
however, in Figure~\ref{fig:mt_ppbar} that the $\Lambda$ data start to have a
slightly different slope above $m_t\sim 2.5$ GeV than the pions, so that it
approaches the kaon data. This does not seem to be the case for the other mesons,
$\eta$'s and kaons, though.

In summary, we showed that the transverse mass spectra of hadrons at RHIC
follows one universal function of $m_t$. Radial flow is not necessary to
describe the data. The same universal behavior seems to present in
proton-antiproton collider data. The spectra at different centralities can be
rescaled into each other up to $m_t\sim 2$ GeV. The scaling parameters follow
the pattern expected from the formation of a gluon-saturated state
(color-glass condensate) if an additional vacuum scale is introduced.

We thank Barbara Jacak, Julia Velkovska, and Nu Xu for many helpful
discussions. JSB thanks RIKEN BNL Research Center and the Nuclear Theory Group
at BNL for their kind hospitality.  This manuscript has been authorized with
the U.S. Department of Energy under Contracts No.\ DE-AC02-98CH10886 and No.\
DE-FG-02-93ER-40764.

\end{document}